\documentclass[10pt,a4paper]{article}

\usepackage[T1]{fontenc}
\usepackage[utf8]{inputenc}
\usepackage{lmodern}
\usepackage{geometry}
\usepackage{hyperref}
\usepackage{graphicx}
\usepackage{verbatim}
\usepackage{amssymb}
\usepackage{tikz,tikz-cd}
\usetikzlibrary{arrows.meta,positioning}
\tikzcdset{every label/.append style = {font = \normalsize}}
\tikzcdset{every diagram/.append style = {row sep=2cm, column sep=2cm}}
\tikzcdset{every arrow/.append style = {outer xsep=1pt}}
\definecolor{fileboxbg}{HTML}{dde8ff}
\definecolor{fileboxborder}{HTML}{6688bb}

\newcommand{\RAG}{RAG}
\newcommand{\LLM}{LLM}
\newcommand{\LaTeXsource}{\LaTeX{} source}
\usepackage{ai-annotation}  

\newcommand{\cp}{\times}
\newcommand{\prjl}{{\ll}}
\newcommand{\prjr}{{\gg}}
\newcommand{\splt}{\mathbin{\vartriangle}}

\title{AI-Friendly \LaTeX:\\
Using \LaTeX{} Code as a Knowledge Source for Retrieval-Augmented Generation}
\author{Tom Verhoeff\thanks{Department of Mathematics \& Computer Science, Eindhoven University of Technology; \texttt{T.Verhoeff@tue.nl}}}
\date{\today}

\begin{document}

\maketitle

\begin{abstract}
Large language models can answer questions about textbooks, lecture notes, and programming exercises more reliably when their answers are grounded in an explicit knowledge source. Retrieval-augmented generation (\RAG) is a common approach: relevant fragments of a document are retrieved and inserted into the model context before answering. For mathematical and technical material, the original \LaTeX{} source can be a better starting point than a PDF, because it contains structural information, labels, sectioning commands, macros, and authorial intent that are often lost or distorted in PDF extraction. However, \LaTeX{} source is not automatically AI-friendly. Cross-references must be resolved, custom macros must be interpreted, exercises and examples must be identified, and author-supplied semantic metadata may be needed. This article describes a focused preprocessing approach for turning \LaTeX{} source, together with its compiled auxiliary files and optional author annotations, into Markdown and JSONL chunks suitable for indexing in a vector database.
\end{abstract}

\section{Introduction}

Textbooks and lecture notes are increasingly used as knowledge sources for AI assistants. A course assistant for beginning programmers, for example, may be expected to answer questions about exercises, compiler errors, and concepts from the textbook while respecting teacher-supplied constraints: do not reveal complete solutions, use the terminology of the course, and refer students to the relevant sections.

A related scenario arises when an AI coding assistant helps a researcher
working on a mathematics paper.
The assistant may need to consult prior theorems, notation conventions,
or formal definitions from related articles and theses for which the
\LaTeX{} source is available.
The requirements are similar but the material is typically denser:
a master's thesis may exceed 100~pages of notation-heavy mathematics
where the structural integrity of the source is especially valuable.

A tempting solution is to give the AI system the PDF of the textbook, article, or thesis. This is convenient, but it is rarely ideal. The PDF is the final visual artifact, not the semantic source. It may obscure structure, damage code formatting, lose labels, split paragraphs across pages, and turn mathematical notation into difficult-to-interpret text. By contrast, the \LaTeXsource{} contains sectioning commands, labels, environments, macro names, and often a modular file structure. These can be valuable for retrieval and grounding.

The central claim of this article is that \LaTeX{} source can serve as a high-quality knowledge source for \RAG{}~\cite{lewis2020rag}, provided that it is preprocessed in a way that respects both its typographical and semantic roles.
The remainder of this article presents an overview of the preprocessing pipeline and then examines each core challenge in turn: cross-references, macros, figures, and author annotations.
Authoring conventions that make \LaTeX{} source more AI-friendly from the start are summarized, followed by related work and a discussion of limitations.
A Python implementation is openly available~\cite{repo}.

\section{The basic problem}

A \RAG{} system typically performs the following steps:

\begin{enumerate}
\item Split a document into chunks.
\item Compute an embedding vector for each chunk.
\item Store the embedding vector together with the chunk text and metadata in a vector database.
\item Embed the user's question.
\item Retrieve the most relevant chunks.
\item Provide those chunks to the \LLM{} as context for answering.
\end{enumerate}

For ordinary prose this can be straightforward. For \LaTeX{} documents it is less so. The source may contain commands such as:

\begin{verbatim}
As explained in Section~\ref{sec:loops}, a while-loop repeats
as long as its guard evaluates to true. See also page~\pageref{sec:loops}.
\end{verbatim}

The literal source text is not what we want to embed. A student is unlikely to ask about \verb|sec:loops|. The RAG representation should instead contain something like:

\begin{verbatim}
As explained in Section 4.3, a while-loop repeats as long as its guard
evaluates to true. See also page 71.
\end{verbatim}

The same issue arises with custom macros. Some macros are merely typographical:

\begin{verbatim}
\newcommand{\code}[1]{\texttt{#1}}
\end{verbatim}

Such a macro can safely be converted to Markdown code notation. Other macros introduce semantic notation:

\begin{verbatim}
\newcommand{\splt}{\mathbin{\triangle}}
\end{verbatim}

Here \verb|\splt| may denote the split combinator from functional programming. Expanding it to a triangle glyph is not necessarily helpful for an AI system. The macro name may be more informative than its visual expansion, provided that the preprocessor also supplies its meaning.

\section{Inputs and outputs of the preprocessor}

The preprocessor accepts the following inputs:

\begin{itemize}
\item the main \LaTeX{} file, for example \verb|book.tex|;
\item the compiled auxiliary file, for example \verb|book.aux|;
\item included \LaTeX{} files, discovered from \verb|\input| and \verb|\include| commands;
\item included auxiliary files, discovered from \verb|\@input| entries in the main \verb|.aux| file;
\item optional author annotations, for example \verb|notation.yaml| or \verb|rag.yaml|.
\end{itemize}
The outputs are:
\begin{itemize}
\item a readable Markdown rendering, such as \verb|book.rag.md|;
\item a label table, such as \verb|labels.json|;
\item a chunk file, such as \verb|chunks.jsonl|, containing one JSON object per chunk.
\end{itemize}
A typical chunk object contains:
\begin{verbatim}
{
  "id": "chunk-00017",
  "heading_path": ["Chapter 4", "While-loops"],
  "source_file": "chapter4.tex",
  "start_line": 120,
  "end_line": 168,
  "markdown": "### While-loops\nA while-loop repeats ...",
  "embedding_text": "Location: Chapter 4 > While-loops\n\n...",
  "metadata": {
    "kind": "explanation",
    "labels": ["sec:while-loops"],
    "page_start": 71,
    "page_end": 74
  }
}
\end{verbatim}

The vector database stores one row per chunk. The embedding vector is computed from \verb|embedding_text|. The human-readable \verb|markdown| and the other fields are stored as metadata or payload. At query time the student's question is embedded, similar chunks are retrieved, and the retrieved Markdown is inserted into the model context together with its metadata.

\section{Why YAML as input and JSON as output?}

The preprocessing pipeline proposed here uses two kinds of structured data: author-written configuration files and machine-generated output files. These have different audiences.

YAML is suitable for author-written configuration. It is relatively readable, supports comments, and allows multi-line text blocks. This matters when an author writes semantic annotations such as explanations of notation, aliases, pedagogical restrictions, or descriptions of exercise roles. For example:

\begin{verbatim}
macros:
  splt:
    name: "split combinator"
    meaning: >
      The split combinator combines two functions with the same input
      into a pair-valued function.
    aliases:
      - split
      - triangle operator
    example_latex: "f \\splt g"
    example_text: "f \\splt g maps x to (f x, g x)."
\end{verbatim}

JSON is better suited for generated output and machine interchange. It is stricter, widely supported, and natural as a line-oriented format such as JSONL, where each line represents one chunk to be embedded and stored. For example:

\begin{verbatim}
{"id":"chunk-00017", "embedding_text":"Location: Chapter 4 > While-loops\n...",
 "metadata":{"source_file":"chapter4.tex", "start_line":120}}
\end{verbatim}

Thus the distinction is not arbitrary. YAML is used where the human author writes and maintains semantic intent; JSON or JSONL is used where software consumes generated data.

\section{The preprocessing pipeline}
\label{sec:pipeline}

The preprocessor runs the following major steps. The subsequent sections address each stage in turn.

\begin{enumerate}
\item Read the main \LaTeX{} file.
\item Recursively follow \verb|\input| and \verb|\include| commands.
\item Read the main \verb|.aux| file and recursively follow \verb|\@input| entries.
\item Parse \verb|\newlabel| entries to build a label table.
\item Load optional YAML annotations.
\item Convert selected LaTeX structures to Markdown.
\item Resolve \verb|\ref|, \verb|\eqref|, and \verb|\pageref|.
\item Interpret known macros according to the annotation registry.
\item Preserve semantic notation and enrich the embedding text.
\item Chunk the result by section, subsection, definition, example, or exercise.
\item Write Markdown, JSON labels, and JSONL chunks.
\end{enumerate}

\begin{figure}[t]
\centering
\begin{tikzpicture}[
    node distance=4mm,
    box/.style={draw, rounded corners=2pt, minimum width=74mm, minimum height=7mm,
                align=center, font=\small, fill=white},
    file/.style={draw=fileboxborder, rounded corners=2pt, minimum width=74mm, minimum height=7mm,
                 align=center, font=\small\ttfamily, fill=fileboxbg},
    arr/.style={->, >=Stealth}
  ]
  \node[file] (in)  {.tex source \quad .aux files};
  \node[box, below=of in]  (s1) {load labels \& read source};
  \node[box, below=of s1]  (s2) {extract figure blocks};
  \node[box, below=of s2]  (s3) {convert \LaTeX{} to Markdown};
  \node[box, below=of s3]  (s4) {chunk \& make figure chunks};
  \node[box, below=of s4]  (s5) {write outputs};
  \node[file, below=of s5] (out) {chunks.jsonl \quad labels.json \quad .rag.md};
  \draw[arr] (in)  -- (s1);
  \draw[arr] (s1)  -- (s2);
  \draw[arr] (s2)  -- (s3);
  \draw[arr] (s3)  -- (s4);
  \draw[arr] (s4)  -- (s5);
  \draw[arr] (s5)  -- (out);
\end{tikzpicture}
\caption{Major stages of the \LaTeX{} \RAG{} preprocessing pipeline.
Input and output files are shown with a gray background.}
\label{fig:pipeline}
\AIDescription{
  A vertical flowchart with seven elements connected by downward arrows.
  The top element (gray) lists the input files: .tex source and .aux files.
  Below it are five white processing boxes: (1) load labels and read source,
  (2) extract figure blocks, (3) convert LaTeX to Markdown,
  (4) chunk and make figure chunks, (5) write outputs.
  The bottom element (gray) lists the output files:
  chunks.jsonl, labels.json, and .rag.md.
}
\end{figure}
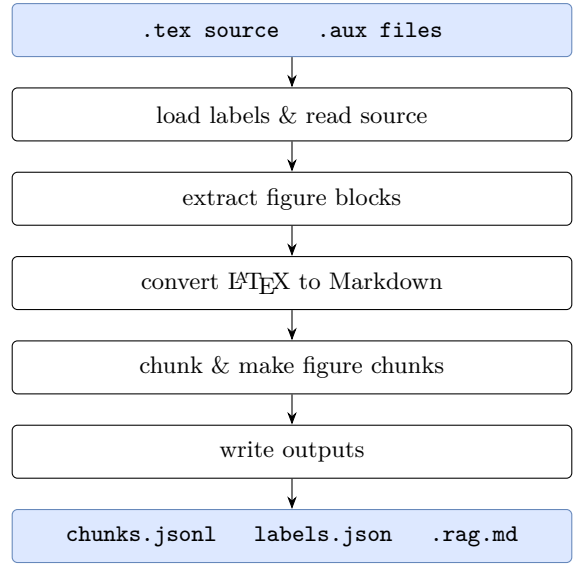

Figure~\ref{fig:pipeline} shows a schematic view of the major pipeline stages.

The preprocessor is conservative: it does not attempt to be a complete \TeX{} interpreter. When it encounters uncertain material, it leaves it in place and emits a warning. For \RAG{}, a robust partial conversion is often more useful than a fragile attempt at perfect expansion.

\section{Resolving references}
\label{sec:references}

After compilation, a LaTeX auxiliary file contains entries such as:

\begin{verbatim}
\newlabel{sec:while-loops}{{4.3}{71}{While-loops}{section.4.3}{}}
\end{verbatim}

The preprocessor parses these entries and builds a label table:

\begin{verbatim}
{
  "sec:while-loops": {
    "ref": "4.3",
    "page": "71",
    "title": "While-loops",
    "anchor": "section.4.3"
  }
}
\end{verbatim}

Then the source fragment:

\begin{verbatim}
See Section~\ref{sec:while-loops} on page~\pageref{sec:while-loops}.
\end{verbatim}

can be converted to:

\begin{verbatim}
See Section 4.3 on page 71.
\end{verbatim}

This simple transformation is already useful. It makes chunks more self-contained and improves retrieval, because the embedded text contains the same surface forms that a student or teacher is likely to use.

When the document uses the \texttt{cleveref} package, the auxiliary file also
contains companion entries:

\begin{verbatim}
\newlabel{sec:while-loops@cref}{{[section][3][4]4.3}{[1][71][]71}}
\end{verbatim}

The first bracketed word, \texttt{section}, is the counter type.
The preprocessor reads these \verb|@cref| entries and resolves
\verb|\cref{sec:while-loops}| to \texttt{section~4.3} and
\verb|\Cref{sec:while-loops}| to \texttt{Section~4.3},
using a built-in table of 25 counter types.

\section{Handling macros}
\label{sec:macros}

Not all macros should be treated alike. We distinguish at least four classes.

\subsection{Typographical macros}

Typographical macros affect presentation but not meaning:

\begin{verbatim}
\newcommand{\code}[1]{\texttt{#1}}
\end{verbatim}

These can be expanded or normalized in Markdown:

\begin{verbatim}
Use `while` rather than `for` in this exercise.
\end{verbatim}

\subsection{Semantic notation macros}

Semantic notation macros name concepts, operators, or conventions. The split combinator is an example:

\begin{verbatim}
\newcommand{\splt}{\mathbin{\vartriangle}}
\end{verbatim}

In the source, one might write:

\begin{verbatim}
The combinator $f \splt g$ maps an argument to a pair.
\end{verbatim}

The Markdown may preserve the notation:

\begin{verbatim}
The combinator $f \splt g$ maps an argument to a pair.
\end{verbatim}

But the embedding text should be enriched:

\begin{verbatim}
The combinator $f \splt g$ maps an argument to a pair.

Notation note: \splt means "split combinator". It combines two
functions with the same input into a pair-valued function.
\end{verbatim}

This is an important principle: the representation shown to students and the representation embedded for retrieval need not be identical.

\subsection{Structural macros}

Some macros encode document structure:

\begin{verbatim}
\Exercise{4.3}{Predict the output of the following program.}
\end{verbatim}

Rather than expanding such macros typographically, the preprocessor interprets them structurally:

\begin{verbatim}
### Exercise 4.3

Predict the output of the following program.
\end{verbatim}

with metadata:

\begin{verbatim}
{"kind":"exercise", "exercise_id":"4.3"}
\end{verbatim}

\subsection{Unknown macros}

Unknown macros should not silently disappear. The preprocessor preserves them and emits a warning. Unknown semantic macros are opportunities for improving the author's annotations.

\section{Handling figures}
\label{sec:figures}

Figures require special treatment in a \RAG{} pipeline. A figure is not merely a visual decoration; it often contains essential semantic information that students may ask about. This applies both to figures generated directly in \LaTeX{}, such as TikZ diagrams, and to externally included images such as PDF, PNG, or JPEG files.

A naive preprocessing strategy might ignore figures entirely or only retain their captions. This is insufficient. Captions typically identify a figure but do not fully explain the knowledge encoded in it. For retrieval purposes, figures need textual surrogates: descriptions that explain what the figure means.

A useful principle is therefore:

\begin{quote}
Figures should be treated as separate knowledge objects.
\end{quote}

The pipeline extracts the following information from each figure:

\begin{itemize}
\item the figure label;
\item the caption;
\item the surrounding section and textual context;
\item the included image filename or TikZ source;
\item the page number;
\item an author-supplied semantic description;
\item optionally, a rendered image representation.
\end{itemize}

For example:

\begin{verbatim}
\begin{figure}
  \centering
  \includegraphics[width=.6\textwidth]{pipeline.pdf}
  \caption{A preprocessing pipeline for LaTeX-based RAG.}
  \label{fig:pipeline}
\end{figure}
\end{verbatim}

may become the following figure chunk:

\begin{verbatim}
{
  "id": "figure-00012",
  "kind": "figure",
  "label": "fig:pipeline",
  "caption": "A preprocessing pipeline for LaTeX-based RAG.",
  "image_file": "pipeline.pdf",
  "embedding_text":
    "Figure: A preprocessing pipeline for LaTeX-based RAG. ...",
  "metadata": {
    "section": "2.3",
    "page": 34
  }
}
\end{verbatim}

The vector database should primarily embed the caption, semantic description, and nearby explanatory text. The raw image file itself is not usually suitable as the primary retrieval representation. Likewise, the raw TikZ source code is often too low-level to serve directly as embedding text.

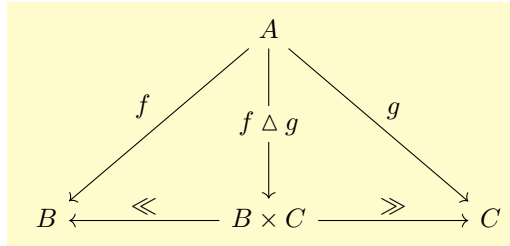
\begin{figure}[hbt]
\centering
\colorbox{yellow!25}
{\setlength{\mathsurround}{0pt}
\begin{tikzcd}[ampersand replacement=\&, background color=yellow!25]
    \& A \ar[dl, swap, "f"] \ar[d, "f \splt g" description] \ar[dr, "g"]
    \& \\
  B \ar[r, leftarrow, "\prjl"]
    \& B \cp C \ar[r, "\prjr"]
    \& C
\end{tikzcd}}
\caption{Commuting diagram for the split combinator.}
\label{fig:split-diagram}
\AIDescription{
  The figure shows two functions f:X to A and g:X to B
  combined into a single function from X to A times B.
}
\end{figure}

TikZ figures deserve particular attention. A TikZ diagram may contain highly meaningful mathematical or categorical structure while remaining difficult to interpret from its source code alone (see Figure~\ref{fig:split-diagram}).

For example:
\begin{verbatim}
\begin{figure}
\centering
\begin{tikzcd}
  ...
\end{tikzcd}
\caption{Commuting diagram for ...}
\label{fig:split-diagram}
\end{figure}
\end{verbatim}
The caption alone is often too brief for retrieval. A better approach is to supply a semantic description explicitly:
\begin{verbatim}
\begin{figure}
...
\caption{Commuting diagram for ...}
\label{fig:split-diagram}
\AIDescription{
  The figure shows two functions f:X to A and g:X to B
  combined into a single function from X to A times B.
}
\end{figure}
\end{verbatim}

The command \verb|\AIDescription| is a no-op during normal compilation so it has no typographic effect.
It is provided by the companion \texttt{ai-annotation} package (see Appendix~\ref{app:author-instructions}):

\begin{verbatim}
\usepackage{ai-annotation}
\end{verbatim}

The preprocessing script recognizes the command and extracts its argument as semantic metadata.

The macro approach is preferred over \LaTeX{} comments for AI-facing annotations; Section~\ref{sec:no-comments} explains in detail why comment-based approaches are unreliable.
The same approach applies to other annotation macros provided by the package: \verb|\AIDeclareNotation| for semantic notation, \verb|\AINote| for inline explanations, and \verb|\AIChunkBreak| for explicit chunk boundaries.

The resulting embedding text for the figure may then become:

\begin{verbatim}
Figure: Commuting diagram for the split combinator.

The figure shows two functions f:X to A and g:X to B
combined into a single function from X to A times B,
from which f and g can be obtained by left and right projections.
\end{verbatim}

This approach is especially important for educational material. Students often refer to figures indirectly:

\begin{quote}
``What does the triangle diagram mean?''

``Why does the arrow split into two arrows here?''

``What does the picture on page 42 show?''
\end{quote}

\begin{figure}[t]
\centering
\includegraphics[width=\textwidth]{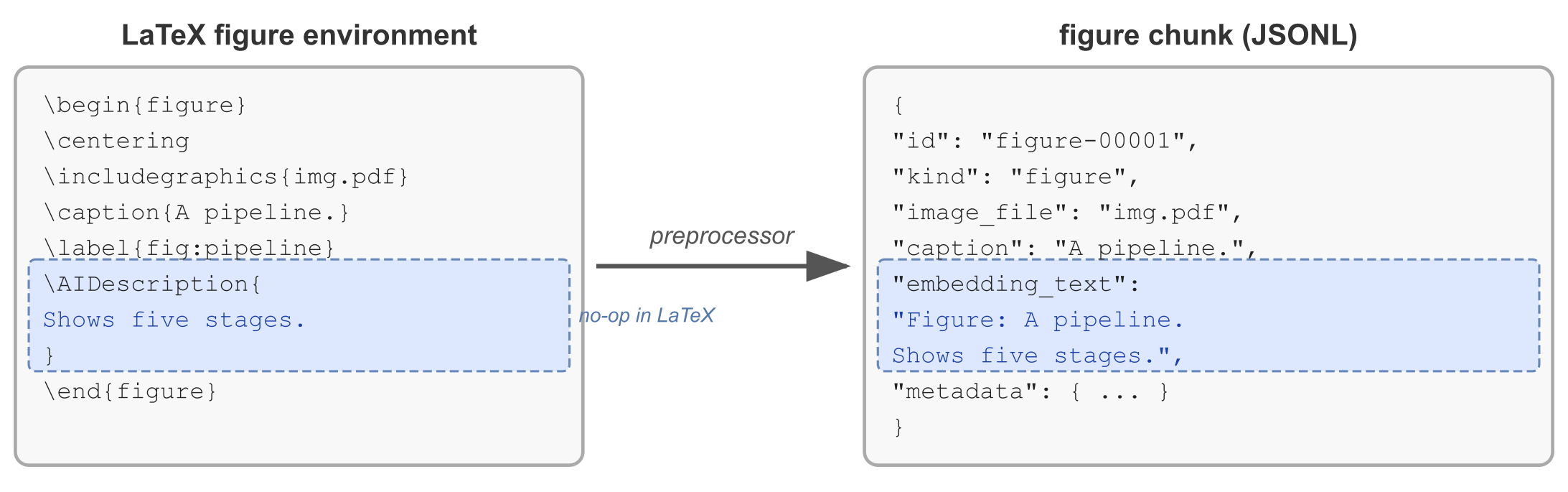}
\caption{Transformation of a \LaTeX{} figure environment (left) into a \RAG{} chunk (right).
The \texttt{\textbackslash{}AIDescription} block (shaded, left) is a no-op during typesetting
but becomes the semantic description in the embedding text (shaded, right).}
\label{fig:figure-chunk}
\AIDescription{
  Two side-by-side boxes connected by an arrow labeled ``preprocessor''.
  The left box shows a LaTeX figure environment containing \textbackslash{}centering,
  \textbackslash{}includegraphics, \textbackslash{}caption, \textbackslash{}label,
  and a highlighted \textbackslash{}AIDescription block with the text
  ``Shows five stages.''.
  The right box shows the resulting JSONL chunk with fields id, kind,
  image\_file, caption, and a highlighted embedding\_text field that reads
  ``Figure: A pipeline.  Shows five stages.''.
}
\end{figure}

Without semantic descriptions, such questions may fail to retrieve the relevant material.
Figure~\ref{fig:figure-chunk} illustrates how a \LaTeX{} figure environment is transformed into a \RAG{} chunk, with the \verb|\AIDescription| content becoming part of the embedding text.

The broader principle is therefore:

\begin{quote}
Figures should have machine-readable semantic descriptions, not only visual renderings.
\end{quote}

This principle applies equally to externally included graphics, TikZ diagrams, commutative diagrams, automata, geometric illustrations, plots, and other forms of visual mathematical communication.

\section{Author annotations}
\label{sec:annotations}

An author can supply semantic metadata to the preprocessor using one of two approaches: in-source declaration macros, or an external YAML file. A third superficially attractive option---tagged \TeX{} comments---does not work and should be avoided.

\subsection{Why tagged comments do not work}
\label{sec:no-comments}

It is tempting to annotate using special comments:

\begin{verbatim}
%AI-MACRO \splt name="split combinator"
\newcommand{\splt}{\mathbin{\triangle}}
\end{verbatim}

Such comments are safe for ordinary \LaTeX{} compilation, but they are invisible to the preprocessor. The preprocessor strips \TeX{} comments during source loading, before any other processing stage (see Section~\ref{sec:pipeline} for the pipeline overview). A comment-based annotation therefore cannot reach the stages that build the macro registry, enrich the embedding text, or generate the notation glossary. This is not an implementation oversight: stripping comments early is correct behavior, because it prevents comment syntax from interfering with macro recognition. In-source annotations must use macro arguments, not comments.

\subsection{Semantic declaration macros}
\label{sec:decl-macros}

The preferred in-source approach uses the \texttt{ai-annotation} package, which provides all annotation macros as no-ops so that ordinary compilation ignores them:

\begin{verbatim}
\usepackage{ai-annotation}
\end{verbatim}

A declaration then accompanies each semantic macro definition:

\begin{verbatim}
\newcommand{\splt}{\mathbin{\vartriangle}}
\AIDeclareNotation
  {\splt}
  {split combinator}
  {Combines two functions with the same input into a pair-valued function.}
\end{verbatim}

For ordinary compilation, \verb|\AIDeclareNotation| does nothing.
For preprocessing, it is a machine-readable declaration of intent: because macro arguments survive comment stripping and are part of the document source tree, they are available to every subsequent processing stage.
The preprocessor scans all source lines for \verb|\AIDeclareNotation| declarations
and merges them with any YAML annotations; YAML takes precedence on conflicts,
making it easy to supplement or override in-source declarations without touching the source.

This approach is well suited to new documents where the author controls the preamble, and to annotations that must be positionally coupled to their content---such as \verb|\AIDescription| inside a figure environment, where the annotation cannot sensibly be expressed without knowing the exact location. The main disadvantage is that annotations are distributed across all \texttt{.tex} files in a multi-file document, making it harder to audit coverage.

\subsection{External YAML file}
\label{sec:yaml-annot}

The alternative is a separate YAML file, for example \verb|book.rag.yaml|:

\begin{verbatim}
macros:
  splt:
    name: "split combinator"
    meaning: >
      Combines two functions with the same input into a pair-valued function.
    aliases:
      - split
      - triangle operator
    example_latex: "f \\splt g"
    example_text: "f \\splt g maps x to (f x, g x)."

pedagogy:
  visibility:
    - environment: solution
      access: teacher_only
\end{verbatim}

The YAML file is preprocessor input only: it is consumed when building the macro registry and applying pedagogy rules, and it does not appear in the vector database or in any output chunk. This distinction matters---the YAML file is an authoring tool, not a \RAG{} artifact.

The \verb|macros| section maps each macro name (without backslash) to its annotation.
The preprocessor uses the registry to append a notation note to the \verb|embedding_text| of every chunk whose Markdown contains the macro, and to generate a single \verb|kind="glossary"| chunk listing all registered macros.
The \verb|pedagogy.visibility| rules cause the preprocessor to inject a \verb|"visibility"| field into the \verb|metadata| of matching structural chunks (e.g.\ \verb|"visibility": "teacher_only"| on \verb|solution| chunks); the RAG application is then responsible for filtering on that field at query time.

A third top-level key, \verb|suppress_macros|, accepts a list of macro names
that should be silenced without annotating:

\begin{verbatim}
suppress_macros:
  - graphicspath
  - usetikzlibrary
  - cite
  - url
\end{verbatim}

Macros in \verb|suppress_macros| are excluded from the glossary and notation
enrichment, and the unknown-macro warning is silenced for them.
The primary use case is preamble and bibliography commands that are not
relevant to retrieval and would otherwise produce spurious warnings.
A secondary use case is silencing macros that the in-source scanner
incorrectly picked up from verbatim code examples
(see Section~\ref{sec:limitations}).

This approach is well suited to existing lecture notes or textbooks where the \texttt{.tex} files should not be modified, and is convenient when annotation is done by someone other than the original author, or when a single auditable view of all semantic metadata is preferred. The main disadvantage is that the YAML file can only reference content that carries a \verb|\label|; positional annotations such as describing a TikZ figure without a label require the in-source macro approach.

\subsection{Choosing between the two approaches}

The two approaches are complementary rather than mutually exclusive. A natural combination is to use \verb|\AIDescription| in figure environments (where position matters) together with a YAML file for the macro registry and pedagogy rules. When starting a new document from scratch, declaration macros throughout the source provide the tightest coupling between annotations and content. When working with existing documents, a YAML file avoids any changes to the \texttt{.tex} source.

\section{Generated notation glossary}
\label{sec:glossary}

When a macro registry is present, the preprocessor generates a notation glossary chunk
(with \verb|kind="glossary"| and \verb|id="glossary-00001"|) listing every registered macro:

\begin{verbatim}
# Notation glossary

## \splt: split combinator

Aliases: split, triangle operator.

The split combinator combines two functions with the same input into a
pair-valued function.

Example: f \splt g maps x to (f x, g x).
\end{verbatim}

This chunk is inserted into the vector database like any other chunk. It helps retrieve explanations when a student asks questions such as ``What is the triangle operator?'' or ``What does split mean here?''

In addition, whenever a registered macro appears in any other chunk, the preprocessor appends a notation note to that chunk's \verb|embedding_text|:

\begin{verbatim}
Notation note: \splt means "split combinator". Combines two functions
with the same input into a pair-valued function. Also known as: split,
triangle operator.
\end{verbatim}

This inline enrichment ensures that even a chunk that only uses the notation---without defining it---is retrievable by the concept name.

\section{Examples}

\subsection{Example: a programming concept}

Original \LaTeX{}:

\begin{verbatim}
\section{While-loops}\label{sec:while-loops}

A \code{while}-loop repeats its body as long as its guard is true.
As explained in Section~\ref{sec:boolean-expressions}, the guard must
be a Boolean expression.
\end{verbatim}

Possible Markdown:

\begin{verbatim}
## While-loops

A `while`-loop repeats its body as long as its guard is true. As explained
in Section 3.1, the guard must be a Boolean expression.
\end{verbatim}

Possible embedding text:

\begin{verbatim}
Location: Chapter 4 > While-loops

A while-loop repeats its body as long as its guard is true. As explained
in Section 3.1, the guard must be a Boolean expression.
\end{verbatim}

\subsection{Example: a functional-programming notation macro}

Original \LaTeX{}:

\begin{verbatim}
\newcommand{\splt}{\mathbin{\triangle}}

The split combinator $f \splt g$ sends an argument $x$ to the pair
$(f\,x,g\,x)$.
\end{verbatim}

Author annotation:

\begin{verbatim}
macros:
  splt:
    kind: semantic_notation
    display: "\\splt"
    name: "split combinator"
    aliases: ["split", "triangle operator"]
    meaning: >
      Combines two functions with the same input into a pair-valued function.
\end{verbatim}

Generated embedding enrichment:

\begin{verbatim}
Notation note: \splt means split combinator. It combines two functions
with the same input into a pair-valued function.
\end{verbatim}

\subsection{Example: exercise and hidden solution}

Original \LaTeX{}:

\begin{verbatim}
\begin{exercise}\label{ex:first-loop}
What is printed by the following program?
\end{exercise}

\begin{solution}
The program prints the numbers 0 through 9.
\end{solution}
\end{verbatim}

For a student-facing tutor, the preprocessor may index the exercise statement but exclude the solution, or store the solution in a separate protected index only available for teacher review. This distinction should be explicit in the metadata:

\begin{verbatim}
{"kind":"exercise", "visibility":"student"}
{"kind":"solution", "visibility":"teacher_only"}
\end{verbatim}

\subsection{Example: textbook reference back to source}

A retrieved chunk should allow the final answer to refer back to the source:

\begin{verbatim}
metadata:
  source_file: chapter4.tex
  start_line: 120
  end_line: 168
  section: "4.3"
  title: "While-loops"
  page_start: 71
  page_end: 74
\end{verbatim}

The AI assistant can then answer:

\begin{quote}
Look again at Section 4.3, especially the explanation of the loop guard on pages 71--72. The key point is that the guard is tested before each iteration.
\end{quote}

\section{AI-friendly LaTeX conventions}

The following conventions make \LaTeX{} source more useful as a knowledge source:

\begin{itemize}
\item Use meaningful labels, such as \verb|sec:while-loops| rather than \verb|s4x3|.
\item Use semantic macros for recurring concepts and notation.
\item Distinguish typographical macros from semantic macros.
\item Mark exercises, examples, definitions, warnings, and solutions structurally.
\item Provide metadata for pedagogical roles, such as student-visible material and teacher-only solutions.
\item Maintain a notation glossary, either by hand or generated from semantic declarations.
\item Avoid relying on purely visual notation when the concept has a name.
\item Preserve code examples in environments that can be recognized reliably.
\end{itemize}

These conventions do not require abandoning ordinary \LaTeX{} practice. Rather, they make authorial intent explicit enough for software to preserve it.

\section{Related work}
\label{sec:related-work}

\paragraph{Retrieval-augmented generation.}
Retrieval-augmented generation was introduced by Lewis et al.\ \cite{lewis2020rag} as a method for grounding language-model outputs in explicit document collections.
Chunking strategy and embedding quality have consistently been identified as key factors in retrieval performance.
The present work addresses a gap at the knowledge-extraction layer: obtaining semantically enriched chunks from structured \LaTeX{} source rather than from rendered PDF or plain text.

\paragraph{\LaTeX{} processing tools.}
LaTeXML \cite{latexml} converts \LaTeX{} source to XML, HTML, and MathML, targeting browser rendering and accessibility.
Pandoc \cite{pandoc} translates between a wide range of document formats, including \LaTeX{} to Markdown or HTML.
Both tools aim at faithful reproduction of the typeset document.
Neither targets the \RAG{} use case: chunking by heading structure, breadcrumb heading paths, resolved cross-references as human-readable text, or embedding-text enrichment from author-supplied semantic annotations.

\paragraph{PDF as a knowledge source.}
The main alternative to processing \LaTeX{} source is to extract text from the compiled PDF\@.
The PDF format was designed for display, not for semantic interchange.
Structure such as section numbers, labels, code blocks, and mathematical notation is frequently lost, damaged, or split across page boundaries.
\LaTeX{} source preserves this structure at the cost of requiring the preprocessing step described in this article.

\section{Limitations and Future Work}
\label{sec:limitations}

The proposed approach is not a replacement for TeX itself. It does not fully expand arbitrary macros, execute conditionals, or reproduce the exact visual document. It also cannot by itself guarantee pedagogical behavior such as never revealing a solution. Such behavior requires additional layers: prompt design, retrieval filtering, answer checking, access control, and evaluation against adversarial student prompts.

There are also copyright and licensing considerations. A textbook may be technically processable but not legally usable in an AI system without permission. This is especially important if student prompts and retrieved excerpts are sent to an external API.

\paragraph{Cleveref.}
Single-label \verb|\cref| and \verb|\Cref| are resolved via the \verb|@cref|
companion entries written by \texttt{cleveref} into the \texttt{.aux} file
(see Section~\ref{sec:references}).
Three gaps remain, in increasing order of difficulty.

\verb|\crefrange{a}{b}|, which produces output such as ``Sections~3--7'',
is straightforward to add once the existing resolution infrastructure is
in place: parse both labels, format each, and join with an en-dash.

Comma-separated lists (\verb|\cref{a,b,c}|) require splitting the argument,
grouping labels by counter type, sorting numerically within each group, and
assembling natural-language output such as ``Sections~2.1 and~3.4''
or ``Figures~3, 5 and~7''.
The logic is well defined but has fiddly edge cases---especially when labels
span multiple counter types (e.g.\ ``Section~3 and Figure~4'')---and
benefits from thorough testing.

Custom noun overrides via \verb|\crefname{counter}{singular}{plural}| require
scanning the \texttt{.tex} preamble for these declarations and merging them
into the noun table at parse time.
The built-in table of~25 counter types handles all standard \LaTeX{} and
\texttt{amsthm} environments; custom environments without a \verb|\crefname|
fall back to a capitalized version of the counter name, which is usually
adequate.
The \verb|suppress_macros| mechanism in the YAML file offers a partial
workaround: users can suppress warnings for unresolved \verb|\cref| calls
without fixing the noun table.

\paragraph{Chunking granularity.}
The pipeline provides two author-control mechanisms for chunk boundaries.
\verb|--min-heading-level| suppresses splits at headings shallower than a
given level while still tracking them in the breadcrumb path.
\verb|\AIChunkBreak| inserts an explicit boundary at any point in the source,
independent of heading structure and the token budget.
These mechanisms address the most common cases but leave finer-grained control
to future work.
Token-budget splitting at paragraph boundaries is a coarse fallback: it cannot
account for semantic coherence, list structure, or the relative importance of
different parts of a long section.
A smarter splitter might, for example, prefer to break before a numbered list
rather than in the middle of one, or keep a theorem and its proof together.
More generally, there is currently no mechanism to mark a passage as
unsplittable or to force it to be kept together with the passage that follows.

\paragraph{Multimodal embeddings.}
The pipeline currently represents figures only through their captions and
\verb|\AIDescription| text.
Models that can embed image content directly could use the source \texttt{.pdf}
or rendered \texttt{.png} to supplement the textual description.

\paragraph{Larger-scale deployment and evaluation.}
The approach has been demonstrated on a single companion document.
Evaluation against real student queries on an actual course textbook---measuring
retrieval precision, recall, and downstream answer quality---would establish
whether the preprocessing choices generalize and where they need refinement.

\section{Conclusion}

\LaTeX{} source is a promising knowledge source for AI systems because it contains structure and intent that are often absent from PDFs. However, it must be preprocessed with care. Cross-references should be resolved, macros should be classified rather than blindly expanded, and semantic annotations should be supplied where typography alone is insufficient. The pipeline uses YAML for human-authored annotations and JSONL for machine-generated chunks. The result is a collection of retrieval units that preserve human-readable notation while enriching the embedded text with the meanings needed for effective retrieval.
Unlike rendering-focused tools such as LaTeXML and Pandoc, the pipeline targets the \RAG{} use case directly: the goal is not faithful typographic reproduction but semantically enriched, retrievable chunks.

A Python implementation of the pipeline described in this article, together with the \LaTeX{} source of this paper, is openly available~\cite{repo}.
Ready-to-run scripts for ingesting chunks into a local vector database and
for searching them from an AI coding assistant are included in the repository
(see Appendix~\ref{app:claude-code}).

The broader goal is AI-friendly \LaTeX: source code that remains suitable for mathematical publishing while also making its semantic and pedagogical structure available to AI systems.

\section*{Acknowledgments}

The author thanks ChatGPT~5.5 (OpenAI) and Claude~Opus~4.7 (Anthropic) for assistance with Python coding and article writing.
The author is the creative driver behind all design decisions and bears sole responsibility for the content.

\appendix

\section{Author Instructions for AI-Friendly \LaTeX}
\label{app:author-instructions}

This appendix provides practical guidance for authors who want their \LaTeX{} source
to work well with the preprocessor.
The instructions below cover the in-source macro approach.
For the external YAML approach and a discussion of when to use each,
see Section~\ref{sec:yaml-annot}.

\subsection{Preamble declarations}

Add the following line to the document preamble.
It loads all AI annotation macros as no-ops that are invisible to the typesetter
but recognized by the preprocessor.

\begin{verbatim}
\usepackage{ai-annotation}
\end{verbatim}

The package defines \verb|\AIDescription|, \verb|\AIDeclareNotation|,
\verb|\AINote|, and \verb|\AIChunkBreak|, all as no-ops during normal compilation.
A \texttt{draft} package option renders them as margin notes and logs them to the
\texttt{.log} file, which can be useful when reviewing annotation coverage.

\subsection{Labels}

Use meaningful, descriptive labels throughout.

\begin{verbatim}
\section{While-loops}\label{sec:while-loops}
\begin{exercise}\label{ex:first-loop}
\begin{figure}...\label{fig:pipeline}
\end{verbatim}

Avoid opaque labels such as \verb|s4x3|, \verb|fig1|, or \verb|eq17|.
Meaningful labels appear verbatim in the label table and in retrieved chunks,
so they benefit both the AI system and human readers of the intermediate files.

\subsection{Figures}

Every figure should have a \verb|\caption| and a \verb|\label|.
Add an \verb|\AIDescription| block after the caption for any figure whose
content is not fully captured by the caption alone:

\begin{verbatim}
\begin{figure}
  \centering
  \includegraphics[width=.6\textwidth]{diagram.pdf}
  \caption{The preprocessing pipeline.}
  \label{fig:pipeline}
  \AIDescription{
    A vertical flowchart with five processing stages connected by downward
    arrows. Inputs at the top: .tex source and .aux files. Processing stages:
    load labels, extract figures, convert to Markdown, chunk, write outputs.
    Outputs at the bottom: chunks.jsonl, labels.json, .rag.md.
  }
\end{figure}
\end{verbatim}

Write the description in plain prose that explains what the figure means or
shows, not merely what it looks like.
Descriptions are especially important for TikZ diagrams, commutative diagrams,
automata, and plots, where the caption alone is typically too brief.

\subsection{Semantic notation macros}

For every custom macro that introduces a named concept or operator, add a
semantic declaration immediately after the definition.
\verb|\AIDeclareNotation| takes three arguments: the macro, its
human-readable name, and a one-sentence explanation.

\begin{verbatim}
\newcommand{\splt}{\mathbin{\vartriangle}}
\AIDeclareNotation
  {\splt}
  {split combinator}
  {Combines two functions with the same input into a pair-valued function.}
\end{verbatim}

The preprocessor scans the source for \verb|\AIDeclareNotation| declarations and uses
them to populate the macro registry.
Both approaches---in-source declarations and the YAML file---are interchangeable for
the macro registry; YAML annotations override in-source ones on conflicts
(see Section~\ref{sec:yaml-annot}).

\subsection{Structural environments}

Mark exercises, solutions, definitions, and similar units with standard
environment names.
The preprocessor recognizes and extracts the following as separate chunks:
\verb|exercise|, \verb|solution|, \verb|definition|, \verb|theorem|,
\verb|lemma|, \verb|proof|, \verb|example|, and \verb|remark|.

\begin{verbatim}
\begin{exercise}\label{ex:first-loop}
What is printed by the following program?
\end{exercise}

\begin{solution}
The program prints the numbers 0 through 9.
\end{solution}
\end{verbatim}

The \verb|kind| field of the resulting chunk (\verb|"exercise"|,
\verb|"solution"|, etc.) can be used to filter retrieval: a student-facing
assistant can exclude \verb|solution| chunks; a teacher-facing assistant can
include them.

\section{Running the Preprocessor and Feeding a RAG System}
\label{app:running}

\subsection{Prerequisites}

The preprocessor has been tested with Python~3.14.
Install it from PyPI:

\begin{verbatim}
pip install latex-rag-preprocessor
\end{verbatim}

Or, within the source repository, use \verb|uv| (no separate installation step):

\begin{verbatim}
uv run python -m latex_rag_preprocessor --help
\end{verbatim}

\subsection{Compile first}

Before running the preprocessor, compile the \LaTeX{} document to produce the
\verb|.aux| file.
Two compilation passes are recommended to settle cross-references:

\begin{verbatim}
pdflatex book.tex
pdflatex book.tex
\end{verbatim}

\subsection{Running the preprocessor}

\begin{verbatim}
latex-rag-preprocessor book.tex --out rag_out
\end{verbatim}

\begin{description}
\item[\texttt{--aux}] Path to the compiled auxiliary file.
  Defaults to the same stem as the \verb|.tex| file.
\item[\texttt{--yaml}] Path to the YAML annotation file (see below).
  Defaults to the same stem as the \verb|.tex| file with a \verb|.rag.yaml|
  extension; silently skipped if the file does not exist.
\item[\texttt{--out}] Output directory (created if absent). Defaults to \verb|rag_out|.
\item[\texttt{--max-tokens}] Approximate token limit per text chunk (default: 900,
  estimated as character count divided by~4).
  Larger values produce fewer, longer chunks; smaller values produce more, shorter ones.
\end{description}

The preprocessor discovers and reads all files reachable through
\verb|\input|, \verb|\include|, and \verb|\@input| automatically;
no additional flags are needed for multi-file documents.

\subsection{YAML annotation file}

Create \verb|book.rag.yaml| alongside \verb|book.tex|, or pass it explicitly
with \verb|--yaml|.
The file has two top-level sections.

The \verb|macros| section maps each macro name (without backslash) to its
annotation. Only \verb|name| is required; all other fields are optional.

\begin{verbatim}
macros:
  splt:
    name: "split combinator"
    meaning: >
      Combines two functions with the same input into a pair-valued function.
    aliases:
      - split
      - triangle operator
    example_latex: "f \\splt g"
    example_text:  "f \\splt g maps x to (f x, g x)."
\end{verbatim}

The \verb|pedagogy| section declares visibility constraints on structural
environments.
The preprocessor injects the \verb|access| value as a \verb|"visibility"|
field in the \verb|metadata| of matching chunks; the RAG application is
responsible for filtering on that field at query time.

\begin{verbatim}
pedagogy:
  visibility:
    - environment: solution
      access: teacher_only
    - environment: proof
      access: teacher_only
\end{verbatim}

\subsection{Outputs}

Three files are written to the output directory.

\begin{description}
\item[\texttt{chunks.jsonl}] One JSON object per line.
  Each object has the fields \verb|id|, \verb|kind|, \verb|heading_path|,
  \verb|source_file|, \verb|start_line|, \verb|end_line|, \verb|labels|,
  \verb|markdown|, \verb|embedding_text|, and \verb|metadata|.
  Text chunks use \verb|kind="text"|; figure chunks use \verb|kind="figure"|;
  structural chunks use the environment name, e.g.\ \verb|kind="exercise"|;
  the notation glossary chunk (if any) uses \verb|kind="glossary"|.
\item[\texttt{labels.json}] A dictionary mapping each \LaTeX{} label to its
  resolved reference number, page, title, and anchor.
\item[\texttt{*.rag.md}] A human-readable Markdown rendering of the processed
  source, useful for inspecting the conversion before indexing.
\end{description}

\subsection{Indexing into a vector database}
\label{app:indexing}

The indexing workflow is vector-database-agnostic:
\begin{enumerate}
\item Read \verb|chunks.jsonl| line by line.
\item For each chunk, send the \verb|embedding_text| field to an embedding model
  to obtain a vector.
\item Store the vector together with the chunk's \verb|id|, \verb|kind|,
  \verb|markdown|, \verb|heading_path|, and remaining metadata in the vector database.
\end{enumerate}

The repository includes a ready-to-run script for ChromaDB~\cite{chromadb}
using local sentence-transformer embeddings (no API key required):

\begin{verbatim}
pip install chromadb sentence-transformers
python3 examples/ingest_chromadb.py rag_out/chunks.jsonl
\end{verbatim}

The script is idempotent: re-running it after preprocessing updates upserts
changed chunks without rebuilding the collection.
An \verb|--openai| flag switches to OpenAI \verb|text-embedding-3-small|
for higher retrieval quality at the cost of an API call per chunk.

At query time, embed the user's question with the same model, retrieve the
most similar chunks by vector distance, and pass their \verb|markdown| text
to the language model as context.
The \verb|kind| field allows retrieval filtering before or after the vector
search: exclude \verb|solution| chunks for student-facing assistants,
or boost \verb|exercise| chunks when the question mentions a specific task.

For production deployments the same chunk structure transfers directly to
Pinecone, Weaviate, pgvector, Qdrant, and similar systems.

\subsection{Using preprocessor output with Claude Code}
\label{app:claude-code}

When using an AI coding assistant such as Claude Code~\cite{claudecode}
while working on a mathematics paper, the assistant can search the
preprocessed document base directly from its shell environment.
This is particularly useful for notation-heavy source material---a
research article or a master's thesis---where PDF extraction would
damage the mathematical structure.

The setup is a one-time operation: run the preprocessor, then ingest.
Once the ChromaDB collection exists, Claude Code can issue queries
from its Bash tool without loading any document into the context window:

\begin{verbatim}
python3 examples/search_chromadb.py "Lyapunov stability" --top-k 3
\end{verbatim}

The script prints the matching chunks with their heading paths, source
locations, and full Markdown text---including \LaTeX{} math notation
preserved from the source---as a single tool-call result that lands
directly in the conversation context.

\paragraph{Location-based browsing.}
Two complementary access modes are available.
\emph{Semantic search} (above) ranks results by embedding similarity and
is effective for conceptual queries even when the source uses different
terminology.
\emph{Location-based browsing} retrieves chunks by their position in the
document hierarchy, using a substring match on the stored \verb|heading_path|:

\begin{verbatim}
python3 examples/search_chromadb.py --location "Chapter 4"
python3 examples/search_chromadb.py --location "Chapter 4 > Stability"
python3 examples/search_chromadb.py "criterion" --location "Chapter 4"
\end{verbatim}

The first form returns all chunks under Chapter~4 in document order---a
structural browse without a semantic query.
The second narrows to a specific section.
The third combines semantic ranking with a location filter.
A \verb|--kind| option further restricts retrieval to a specific chunk
type, for example \verb|--kind theorem|.
No re-ingestion is required to use \verb|--location|: the
\verb|heading_path| field is already stored as metadata by the ingest script.

\paragraph{Choosing between search and direct reading.}
For small documents---up to roughly 30\,000 tokens---it may be simpler
to give Claude Code the \verb|*.rag.md| Markdown rendering directly.
The rendering is self-contained, has cross-references resolved to numbers,
and preserves \LaTeX{} math notation faithfully.
The ChromaDB approach scales to larger corpora (a full thesis, a
collection of related papers) and avoids saturating the context window
when only a few relevant passages are needed.

\end{document}